\documentclass[12pt,preprint]{aastex}

\usepackage{psfig}

\begin{document}


\title{Resolving the Innermost Region of the Accretion Disk of the Lensed Quasar Q~2237+0305 through Gravitational Microlensing}



\author{E. MEDIAVILLA\altaffilmark{1,2}, J. JIM\'ENEZ-VICENTE\altaffilmark{3,4}, J. A. MU\~NOZ\altaffilmark{5,6}, T. MEDIAVILLA\altaffilmark{7}}

\altaffiltext{1}{Instituto de Astrof\'{\i}sica de Canarias, V\'{\i}a L\'actea S/N, La Laguna 38200, Tenerife, Spain}
\altaffiltext{2}{Departamento de Astrof\'{\i}sica, Universidad de la Laguna, La Laguna 38200, Tenerife, Spain}
\altaffiltext{3}{Departamento de F\'{\i}sica Te\'orica y del Cosmos, Universidad de Granada, Campus de Fuentenueva, 18071 Granada, Spain}
\altaffiltext{4}{Instituto Carlos I de F\'{\i}sica Te\'orica y Computacional, Universidad de Granada, 18071 Granada, Spain}
\altaffiltext{5}{Departamento de Astronom\'{\i}a y Astrof\'{\i}sica, Universidad de Valencia, 46100 Burjassot, Valencia, Spain.}
\altaffiltext{6}{Observatorio Astron\'omico, Universidad de Valencia, E-46980 Paterna, Valencia, Spain}        
\altaffiltext{7}{Departamento de Estad\'{\i}stica e Investigaci\'on Operativa, Universidad de C\'adiz, Avda Ram\'on Puyol s/n, 11202, Algeciras, C\'adiz, Spain}

\begin{abstract}

We study three high magnification microlensing events, generally recognized as probable caustic crossings, in the optical light 
curves of the multiply imaged quasar Q~2237+0305.  
We model the light curve of each event as the convolution of a standard thin disk luminosity profile with a straight fold caustic. We 
also allow for a linear gradient that can account for an additional varying background effect of microlensing. 
This model not only matches noticeably well the global shape of each of the three independent microlensing events but also gives
remarkably similar estimates for the disk size parameter. The measured average half-light radius, $R_{1/2}=(3.0\pm 1.5)\sqrt{M/0.3M\odot}$
light-days, agrees with previous estimates. 
In the three events, the core of the magnification profile exhibits ``fine structure" related to the innermost region of the accretion 
disk (located at a radial distance of $2.7\pm 1.4$ Schwarzschild radii according to our measurement). Relativistic beaming at the
internal rim of the  accretion disk can explain the shape and size of the fine structure, although alternative 
explanations are also possible. This is the first direct measurement of the size of a structure, likely the innermost stable circular orbit, at $\sim 3$ Schwarzschild radii  in a quasar accretion disk. The monitoring of thousands of lensed quasars with future telescopes will allow the study of the event horizon environment of black holes in hundreds of quasars in a wide range of redshifts $(0.5<z<5)$.

\end{abstract}


\section{Introduction \label{intro}}

The structure of the central engine of quasars can potentially be resolved right to the supermassive black-hole event
horizon by microlensing caustic crossing, a phenomenon related to  gravitational lensing of distant quasars by 
intervening lens galaxies (Chang \& Refsdal 1979, 1984; Wambsganss 2006). The main contributor to the gravitational
field of the lens galaxy is the smoothly distributed dark matter. But there are also compact sources (stars and their 
remnants) that locally break the smoothness of the gravitational field and split each image of the quasar in several
micro images.  When, due to the movement relative to the galaxy, a source changes its  position relative to the 
distribution of stars, the number of micro images can change. Regions where the source is mapped to a different 
number of images are separated by caustic curves where the lens mapping bifurcates and the magnification is 
formally infinity. When a sufficiently small sized source crosses a caustic its magnification rises with a very 
large gradient at the caustic and then decreases following a $D^{-1/2}$ dependence with the distance to the caustic  
(straight fold caustic approximation, Schneider \& Weiss 1987). In this case, the convolution of the source 
luminosity and the caustic magnification profiles can keep a great deal of information about the structure of the source. 
When the size of the source is relatively large, however, the phenomenology of caustic crossing becomes more complex 
and the source structure can be hard to deconvolve ({Shalyapin et al. 2002,} Mosquera, Mu\~noz \& Mediavilla 2009).

Q 2237+0305 is a particularly interesting system for studying caustic crossings. The unusually {low redshift} of the lens galaxy
{allows} the quasar images {to} be seen through the galaxy bulge in a region of high density of stars (and hence of caustics). 
The relatively high rate of caustic crossings is also favored by the high relative velocities between the source and the lens
galaxy. Our aim is to use the caustic crossings observed in the optical light curves available for Q~2237+0305 to study the
source luminosity profile.  {In particular, we would like to test the classical thin disk model,} infer the disk size, and 
set the spatial scale of the possible substructure of the source. Several of these objectives have, in fact, been attempted in 
an ample number of works with heterogeneous methodology and results. A group of authors (Shalyapin et al. 2002, Goicoechea et 
al. 2003,  Koptelova, et al. 2007, Abolmasov \& Shakura 2012) has focused on the region around the High Magnification Event  (HME) peak.
This approach is prone to give small estimates of the disk size. {Another} group of authors has attempted the fit of
the light curve in a region around the event sufficiently large  (some authors considered even the fit of the whole light 
curves) as to constrain the baseline far enough from the caustic crossing event, and also to account for the intrinsic
source variability (Kochanek 2004, Eigenbrod et al. 2008, Anguita et al. 2008, Sluse et al. 2011, Poindexter \& Kochanek 2010). 
The estimates of the different authors for the size of the source exhibit a large scatter (between 0.2 and 10 light-days)
obviously affected by the difference in methodology but also, in many cases, by the lack of information about the 
effective transverse velocity between the source and the galaxy (see, e.g., Table 4 of Sluse et al. 2011).

Very recently we have obtained, from the statistics of the number of caustic crossings, a measurement of the transverse 
velocity in Q~2237+0305 (Mediavilla et al. 2015)  independent of the size of the source that will allow us to map the time 
scale of the light curves to lengths.
In Mediavilla et al. (2015) a preliminary comparative study of the shape of the microlensing events was done to obtain a 
rough estimate of the size of the events from the FWHM of their average profile. However, no direct attempt to fit the 
luminosity profile of the source was {made}.  Our approach in this paper, then, is to analyze the three events 
generally recognized as caustic crossings, removing intrinsic variability. In order to do so, we consider a large 
enough region to constrain the baseline of each caustic crossing event. Finally, we use our independent estimate 
of the velocity to set the spatial scale at the quasar. 

The phenomenology of these three events is not so simple as to allow a good fitting using the straight fold 
caustic model alone. A more comprehensive description of microlensing effects based {on} the numerical 
simulation of microlensing light curves from magnification maps has been reasonably successful in analyzing  
the combined influence of microlensing and {intrinsic} variabillty in several images  at once (Kochanek
2004, Poindexter \& Kochanek 2010).  However, this numerical approach is a computational challenge when 
detailed source models with several parameters (inner and outer radii, scale parameter, inclination, orientation 
respect to the caustic, slope of the luminosity profile, etc.) are to be considered. Thus, we would like to
identify the simplest microlensing phenomenology beyond the straight fold caustic crossing that can be
incorporated into this approximation to obtain good analytical fits. If the complexity of the events arises 
from the combination of the effects of more than one caustic crossing or to large effects of caustic
curvature, the attempt will be very difficult. On the contrary, if the additional complexity corresponds 
to a  background gradient in the magnification, 
we can try to account for this effect by including an ad hoc extra term, linear in the distance to the caustic.
We will show that this last possibility works very well in the case of the three events selected as the best candidates for  single caustic crossing.

Thus, in \S 2 we present and discuss the results of the analysis of the three events under the straight fold caustic 
approximation with an additional linear term. In \S 3 we fit a model including relativistic effects to the fine 
structure detected in the luminosity profiles and determine its spatial scale. 
Finally, in \S 4 the main results are summarized.

\section{Analytical Fitting of three HMEs\label{fitting}}

We start by assuming that an HME can be modeled by a single caustic crossing plus
 a linearly varying background effect of microlensing. To remove the intrinsic
variability of the source we divide the light curve with the HME by the light curve of the image with the smoothest
variations in the epoch of interest.\footnote{Notice that the time delay between images is small enough to be neglected.} This last image could be also affected by microlensing but we will consider that this variation is also included in the linear gradient term.

Thus, we fit each HME with a simple model resulting from the convolution of a straight fold caustic magnification
profile (Schneider \& Weiss 1987, Shalyapin et al. 2002) with a thin disk luminosity profile (Shakura \& Sunyaev 1973) plus a linear term:
\begin{equation}
\label{eq00}
f= c_0+c_J J((t-t_0)/\Delta t)+c_1t ,
\end{equation}
with,
\begin{eqnarray}
J((t-t_0)/\Delta t)= {1\over\pi\Gamma[{11\over 3}]Z[{8\over 3}]}\int_{0}^{+\infty}dx\,x^{-1/2}\int_{-\infty}^{+\infty}dy \left[\exp\left(\left(\sqrt{(x-(t-t_0)/\Delta t)^2+y^2}\right)^{4/3}\right)-1\right]^{-1},\cr\nonumber\\
\end{eqnarray}
where the relevant physical parameters are the disk scale parameter expressed in days, $\Delta t\ =R_\lambda/v_{eff}$, 
and the caustic crossing time,  $t_0$. $R_\lambda$ is the disk scale parameter and $v_{eff}$ the relative velocity
between the caustic and the source. This model is linear in $c_0$, $c_J$ and $c_1$ but non-linear in $t_0$ and $\Delta t$.

We fit the model described in Eq. \ref{eq00} to the three HME generally classified as probable single caustic crossings:
two events in image A at JD (245)1500 (Wyithe et al. 2000, Shalyapin 2001, Goicoechea et al. 2003, Kochanek 2004, 
Gil-Merino et al. 2006, Koptelova et al. 2007, Abolmasov \& Shakura 2012) and at JD 4000 (Eigenbrod et al. 2008a, 2008b), 
and one event in image C at JD 1360 (Wyithe et al. 2000, Shalyapin 2001, Kochanek 2004,  Koptelova et al. 2007, 
Anguita et al. 2008, Abolmasov \& Shakura 2012). We have used the Optical Gravitational Lensing Experiment 
(OGLE, Wozniak et al. 2000, Udalski et al. 2006) optical light-curves of Q~2237+0305 joined, in the case of 
the A image event at JD 1500, with the GLITP (Gravitational Lensing International Time Project) light-curve
(Alcalde et al. 2002). We have selected a large time interval ($\sim 800$ days) around the peak of each 
HME to properly define the caustic baseline, a very important step to perform the comparison of the events. 
To remove intrinsic variability we have divided the light curve of each event by the light curve of the 
image that exhibits less variability during the event (image B for the events A/JD 1500 and C/JD 1360 and 
image D for the event A/JD 4000). 

The model succeeds in fitting the global structure of the three caustic crossings (Figure 1) with
a remarkably good agreement between the scale parameters (Table 1).  To estimate the errors in the 
fluxes  (see {parameter $\langle\sigma_{adj}\rangle$ in} Table 1)  including all the sources of
uncertainty we have computed the average of the absolute value of the differences between points
adjacent in the light curves separated by 2 days or less (a very conservative estimate of the error).
In Figure 2 we have over-plotted the three events after subtraction of the linear component, re-centering, 
and normalization in both amplitude and scale parameter. There is an outstanding global coincidence between 
the profiles very difficult to explain without accepting the hypothesis of a shared common phenomenology. 
Perhaps the coincidence is not so surprising taking into account that the three HME have, in fact, been
selected among others as the best candidates for single caustic crossing.

The average of the scale parameter expressed in days is $\Delta t=(R_\lambda/v_{eff})=56.5$ days, and its dispersion,
$\sigma_{\Delta t}=4.4$ days, corresponding to a half-light radius of $R_{1/2}={2.44} R_\lambda={137.86}\pm {10.74}$ days. 
To map this time interval in the light curve into a length at the quasar plane we will use the measurement of the 
source velocity in Q~2237+0305 by Mediavilla et al. (2015), $v_{eff}=(493\pm246)\sqrt{M/0.17 M_\odot}\,\rm km\,s^{-1}$, consistent with 
previous upper (Wyithe et al. 1999, Gil-Merino et al. 2005) and lower (Poindexter \& Kochanek 2010) limits,  
obtaining, $R_{1/2}=(3.0\pm1.5)\sqrt{M/0.3 M_\odot}$ light-days at $\lambda_{rest}=2018\,\rm\AA$ (notice that 
the main source of uncertainty is $v_{eff}$). This value is in agreement with previous determinations (see 
Mediavilla et al. 2015, Mu{\~n}oz et al. 2015 and references therein).

\section{Resolving the Innermost Region of the Accretion Disk from the Fine Structure of the Magnification Profiles}

As commented in \S \ref{intro}, the possibility of resolving the inner structure of the quasar accretion disk 
just to a few gravitational radii ($R_g=GM/c^2$) is perhaps the most challenging motivation for caustic crossing studies. 
Among the three HME that we are studying, the event at A/JD 1500 is the one with the best available light curves.
In 1999 and 2000 we conducted a daily photometric monitoring (GLITP, Alcalde et al., 2002) of Q~2237+0305 with the 
Nordic Optical Telescope that allowed us to map with good sampling and accuracy the peak of this HME. Its light curve 
does not have a smooth shape at the maximum but shows, in both R and V bands, 
fine structure in the form of a narrow peak followed by a dip and a possible smooth secondary peak 
(two-peaked structure), qualitatively interpreted by Abolmasov \& Shakura (2012) as a signature of 
the inner disk structure. Abolmasov \& Shakura fit the two-peaked structure with a high inclination 
Kerr model that, in principle, could help to estimate sizes in the source, but the required values for 
the black-hole mass and the effective velocity of the source were so different from expectations that no 
consistent estimate of the size of the substructure can be based {on} this model alone.

We will use the independent (not based {on} light curve fitting) measurement of $v_{eff}$ by Mediavilla et 
al. (2015) as an additional constraint to infer the substructure size from the three HME light curves. To do 
that the first step is to look for fine structure (consistent with that found in the event at A/JD 1500)
around the maximum in the events at A/JD 4000 and C/JD 1360 that could confirm or deny that the inner 
region of the accretion disk is resolved in the three caustic crossing magnification profiles.

As seen in \S \ref{fitting} the global shape of the three HME can be consistently fitted by the caustic
crossing model (see Figure 2). However, there are some features of the HME profiles that the model is
unable to fit. Some of them,  like the ones around 400 days or -200 days in Figure 2 are likely related
to the complexity of the HME beyond the straight caustic crossing approximation. 
Around the maximum of the HME, however, the inability of the model to fit the data is more 
likely related to the presence of substructure in the source. If the substructure is real, 
evidence of it  should appear in the three events. To ascertain this, we can look at a zoom 
of the over-plotted profiles in the region around the maximum (see inset in Figure 2). We see 
that the two-peaked structure first detected in the A/JD 1500 event is also present in the
other HME of image A at JD 4000. The event in image C also matches very well the highest peak 
just to the central dip but it does not reach the secondary maximum\footnote{Although this
difference could be explained by noise (the light curve of image C has only a couple of 
points in the region of the secondary peak) the convolution of a Kerr model of high inclination 
with caustics at different position angles can qualitatively explain the existence of either 
one or two narrow peaks in the magnification profile (Abolmasov \& Shakura 2012).}.

To study the significance of the substructure we have averaged the data in 5-day intervals obtaining the mean profile plotted in Figure 3. 
Taking the standard errors in the mean inferred from the averaging as reference, we can
conclude the existence of fine structure in the core of the HME, with a narrow outstanding
peak and a probable secondary smooth maximum. 
As commented above, to measure the size of the fine structure, we use the estimate of 
$v_{eff}$ in Q~2237+0305  by Mediavilla et al. (2015). We find that the time separation
between the peaks of the fine structure (about $50$ days) corresponds to a distance of
$0.8\pm 0.4$ light-days at the quasar that for a black hole mass of $1.2\times 10^9\, M_\odot$ 
(Assef et al. 2011)  corresponds to $5.3\pm 2.7$ Schwarzshild radii, i.e., comparable to the 
diameter of the Innermost Stable Circular Orbit (ISCO) in Schwarzschild space. Alternatively, 
taking as baseline the dip between peaks, the primary peak has a FWHM of $25$ days, 
approximately, that corresponds to $2.7\pm1.4$  Schwarzshild radii. These are strong 
quantitative evidence that the observations of the three selected HME of Q~2237+0305 
are resolving the innermost structure of the quasar accretion disk.

A non-relativistic thin disk model with an inner hole limited by the ISCO may induce 
fine structure at the maximum but  would make less prominent the peak closer to the 
sharp side of the caustic, contrary to the observations. It has been proposed that 
differential relativistic beaming of the approaching and receding sides of the inner
rim of the disk can explain the relative intensity of the peaks (Abolmasov \& Shakura 
2012).  We explore this possibility introducing relativistic beaming\footnote{At
each point of the disk, we correct the flux considering the specific intensity invariant,
$I_\nu/\nu^3$,  and the Doppler and gravitational shifts of the frequency, $\nu$.}
in the thin disk model and considering the convolution of an inclined disk with the 
caustic but using our new measurements of the disk parameter scale and of the size of the fine structure.
Fixing the scale parameter  to the average of the values in Table 1 ($\Delta t=56.5 \rm\, days$) 
and setting the ISCO radius to 3 Schwarzschild radii (in agreement with our estimate of
$2.7\pm1.4$ Schwarzschild radii)  we have fitted the relativistic model to the innermost
region obtaining the fit shown in Figure 3. 
According to this fit, a model with the ISCO at 3 Schwarzschild radii can consistently 
account for the shape of the fine structure when high inclination, $i\sim 73^\circ$, is 
allowed, which can be controversial (Poindexter \& Kochanek 2010, Abolmasov \& Shakura 2012). 
The inclusion of relativistic beaming has a negative impact in the fit of the wings of 
the HME even when changes in the scale parameter or in the slope of the thin disk luminosity
profile are allowed. It is probable that these problems can be addressed by modifying the model 
and, in any case,  the picture that emerges from the relativistic fitting is indeed very 
suggestive, but alternative possibilities such as the presence of an inhomogeneity in the disk 
(bright spot) close to the ISCO should also be considered (see, e.g., Moriyama and Mineshige, 2015).

\section{Conclusions}

The inclusion of a linear term in the analytical model of straight fold caustic crossing by a
thin disk, has allowed us to obtain very good fits to the global shape of three selected HME in the 
light curves of the lensed quasar Q~2237+0305. We found  excellent agreement between the 
independent source half-light radii estimates of the three fits and an outstanding coincidence
between the experimental profiles of the three HME after removing the linear term (in support of 
the adopted hypothesis of straight fold caustic crossing). The most important results of this study are:

1 - We determine an average half-light radius for the source of $R_{1/2}=(3.0\pm1.5)\sqrt{M/0.3 M_\odot}$ light-days
at $\lambda_{rest}=2018\,\rm\AA$ in agreement with other results (see e.g. Mu{\~n}oz et al. 2015 and references therein).

2 - We detect a two-peaked structure in the A/JD 4000 HME that matches reasonably well with the 
analogous structure previously found in the A/JD 1500 HME (see Abolmasov \& Shakura, 2012) and
partially with the narrow peak observed in the C/JD 1360 HME. This coincidence supports the 
detection of fine structure in the HME profiles that can be related to the innermost regions of the accretion disk.   

3 - We set the spatial scale in the HME profiles measuring a separation between 
the peaks of the substructure of about $0.8\pm {0.4}$ light-days equivalent to $5.3\pm {2.7}$ 
Schwarzschild radii for a $1.2\times10^9 M_\odot$ Black Hole. This is  quantitative 
evidence that the innermost region of the accretion disk, at the ISCO for a 
Schwarzshild metric, is resolved in the HME of Q~2237+0305. In fact, relativistic effects, 
like beaming acting on an inclined disk, may explain the intensity ratio between the 
substructure peaks (see also Abolmasov \& Shakura, 2012).

Taking into account the results shown here and the high observational expectations that
some relatively easy experimental improvements  can bring (large telescope, adaptative
optics, multi-band photometry, continued daily and occasionally hourly observations,
etc.), the monitoring of Q~2237+0305 should be resumed as a unique source of information about the black hole environment in quasars. 
In the future, ten years of monitoring of thousands of lensed quasars with the Large Synoptic 
Survey Telescope (Marshall et al. 2010), will provide hundreds of HME (assuming an average
Einstein crossing timescale of 20 years [Mosquera \& Kochanek 2011] and 1 caustic per crossing) 
scanning the innermost environment of supermassive black holes in a sample of quasars with a 
large range in redshifts, $0.5<z<5$ (Mosquera \& Kochanek 2011), greatly improving statistical 
and evolutive studies of  supermassive black holes and accretion disks in quasars.

\acknowledgements{We appreciate the valuable suggestions of the referee. We thank OGLE and GLITP collaborations for  providing  data. This work is supported by  Spanish MINECO,  Junta de Andaluc\'\i a and  Generalitat Valenciana through grants: AYA2010-21741-C03-02, AYA2011-24728, 
AYA2013-47744-C3-1, AYA2013-47744-C3-3-P, AYA2014-53506-P,  FQM-108, and PROMETEO/2014/60.}

\clearpage

\begin{figure}[h]
\plotone{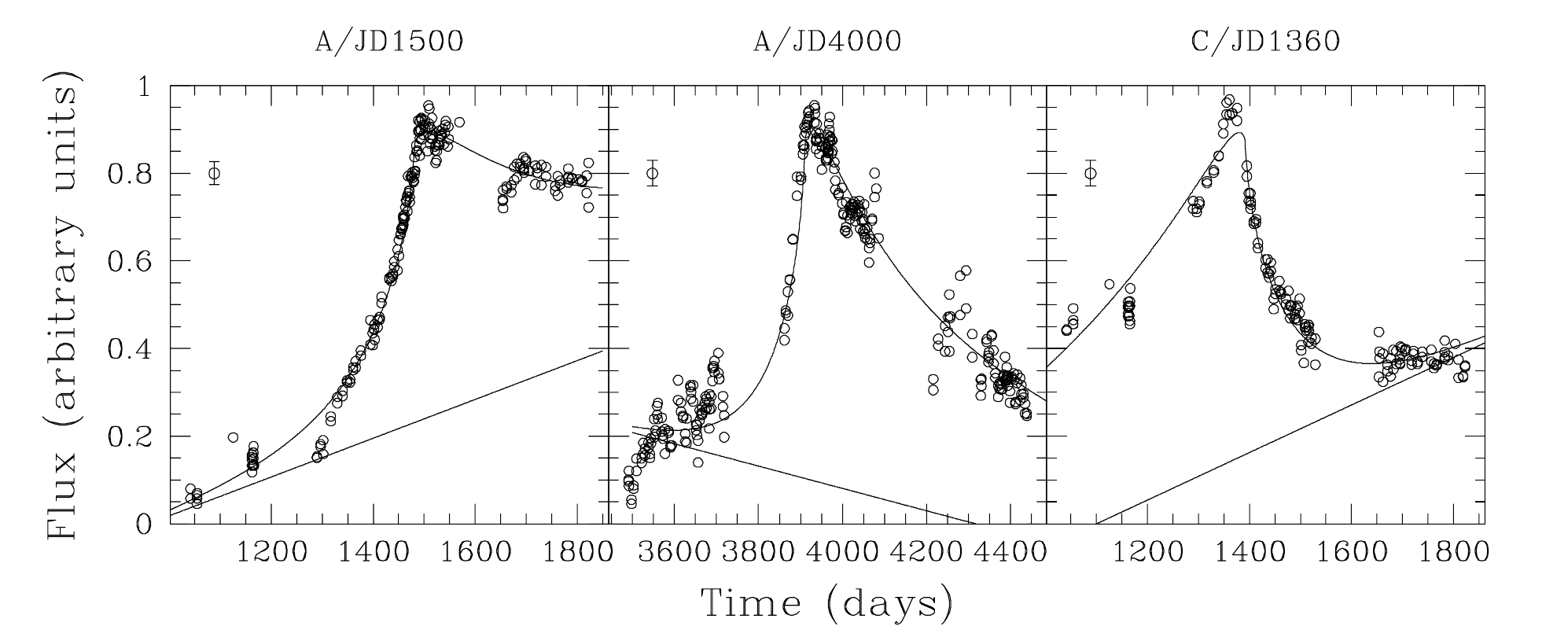}
\caption{Independent fits for each of the three HMEs. Points correspond to the  light curves.
Continuous curves are the fits to the model obtained from the convolution of a thin disk profile
with a straight fold caustic plus a linear term.  Errors bars (see text) are included at the upper left corners.}
\end{figure}

\begin{figure}[h]
\plotone{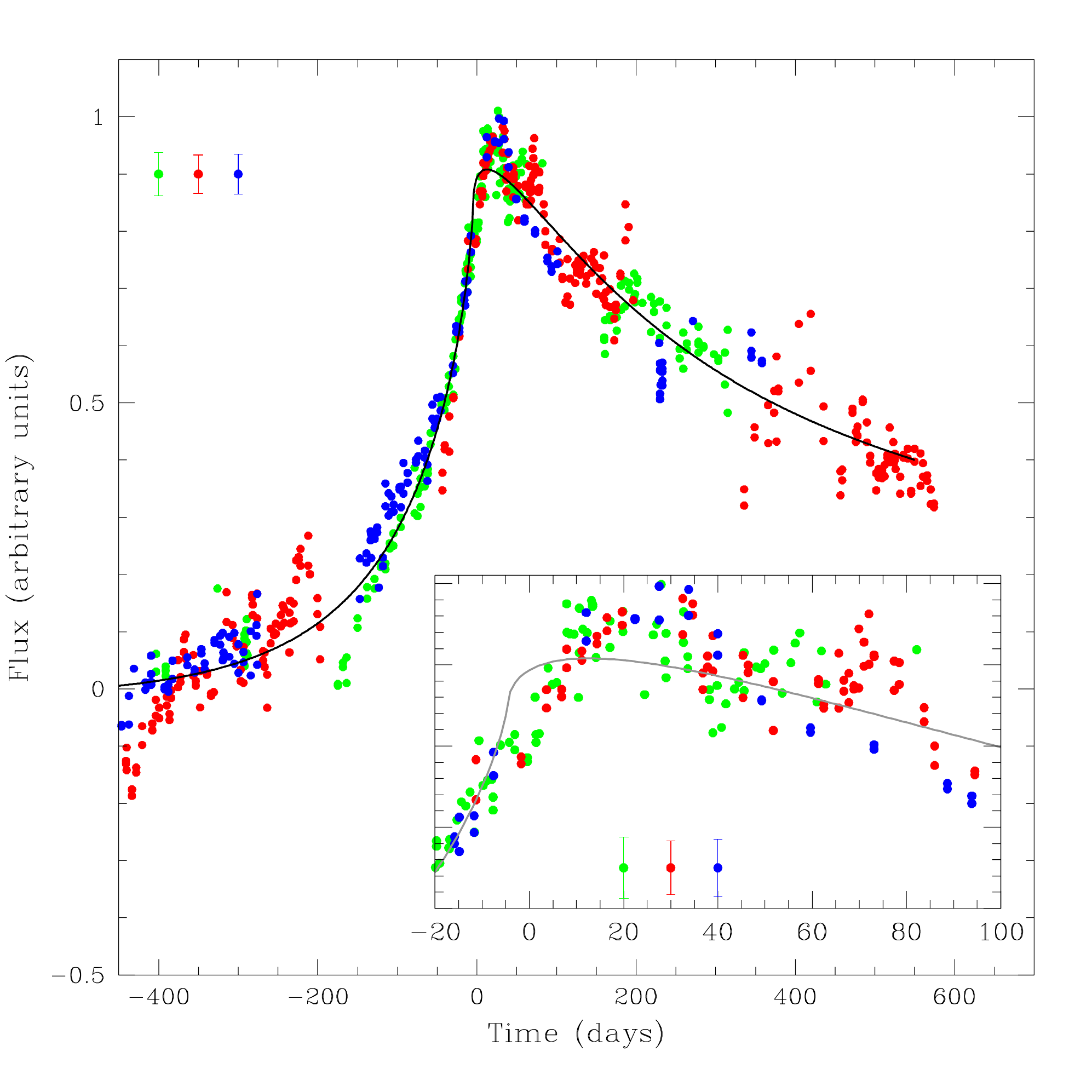}
\caption{Global comparison of the three HMEs and analytical fit to a single caustic crossing.
The inset is a zoom in the region of the maximum.  Dots correspond to the experimental light
curves of each of the three events (green: A/JD 1500, red: A/JD 4000, blue: C/JD 1360), 
after removing the linear terms of the fits (see Figure 1), re-centering, and normalization.
The continuous curve is the fit of a model without linear term to the whole set of data. 
Average error estimates (see {text}) for each  of the HME data are included. \label{fig2}}
\end{figure}

\begin{figure}[h]
\plotone{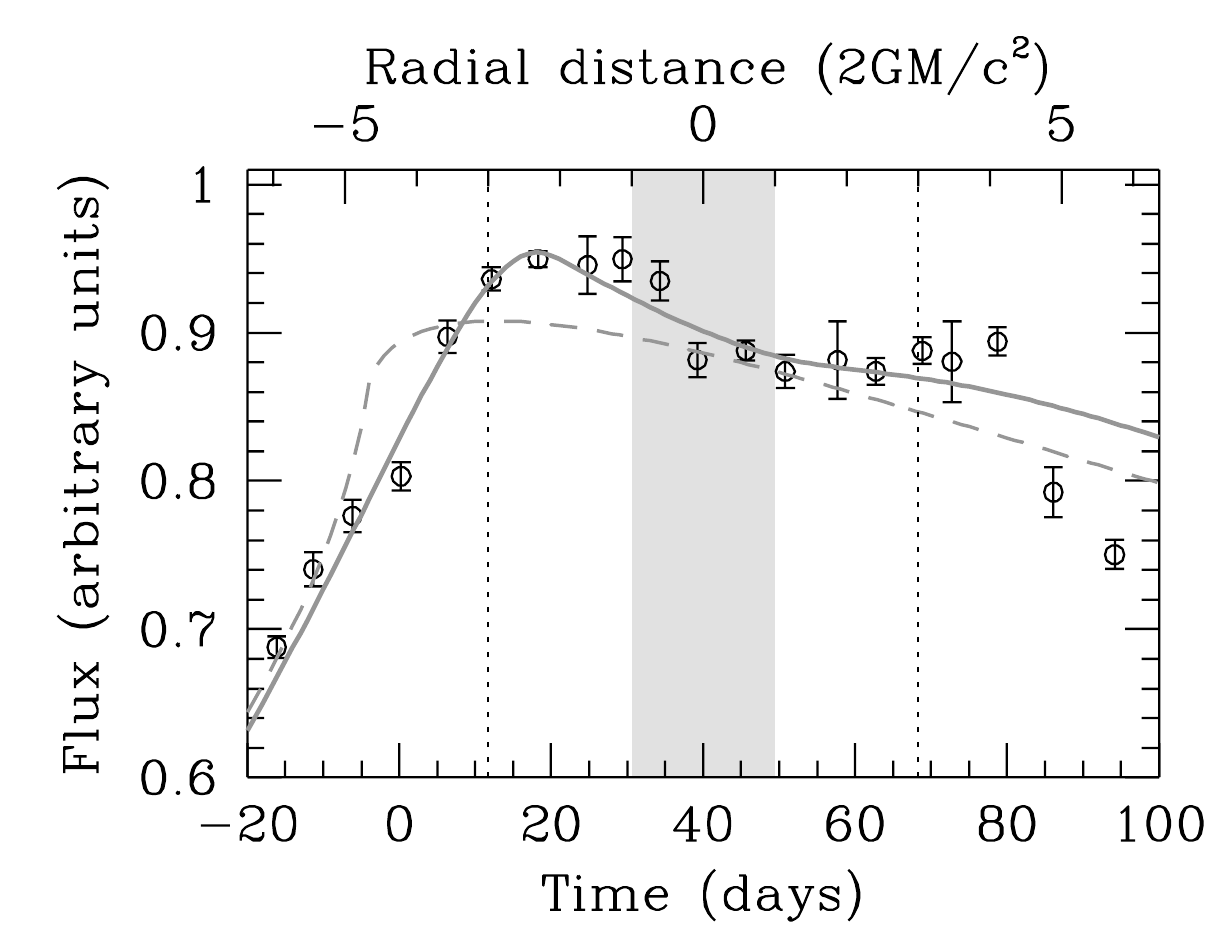}
\caption{Average of the three HMEs of microlensing in the region of the maximum. 
Points correspond to the average in intervals of 5 days. Error bars are the standard 
errors in the mean corresponding to each 5-day interval. The continuous curve corresponds 
to a relativistic model with innermost stable orbit at 3 Schwarzschild radii. The dashed 
curve is the non-relativistic fit of Figure 2. Dotted vertical lines correspond to $\pm 3$ 
Schwarzschild radii (ISCO) and the shaded region to $\pm 1$ Schwarzschild radii (event horizon).}
\end{figure}

\clearpage

\begin{table}
\caption{Fit Parameters for the three HME.}
\medskip
\begin{tabular}{cccc}
\hline
Event & $\Delta t (days)$ & $\langle \sigma_{adj} \rangle$ & $\chi^2_{red}$ \\
\hline
A / JD 1500& $51.5^{+3}_{-3}$ & 0.06 & 1.52 \\
A / JD 4000 & $60.0^{+9}_{-8}$ & 0.02& 3.6 \\
C / JD 1360& $58.0^{+5}_{-5}$ & 0.05 &1.82 \\
\hline
\end{tabular}
\end{table}


\begin{thebibliography}{}

\bibitem[Abolmasov \& Shakura(2012)]{2012MNRAS.423..676A} Abolmasov, P., \& Shakura, N.~I.\ 2012, \mnras, 423, 676 
\bibitem[Alcalde et al.(2002)]{2002ApJ...572..729A} Alcalde, D., Mediavilla, E., Moreau, O., et al.\ 2002, \apj, 572, 729 
\bibitem[Anguita et al.(2008)]{2008A&A...480..327A} Anguita, T., Schmidt, R.~W., Turner, E.~L., et al.\ 2008, \aap, 480, 327 
\bibitem[Assef et al.(2011)]{2011ApJ...742...93A} Assef, R.~J., Denney, K.~D., Kochanek, C.~S., et al.\ 2011, \apj, 742, 93 
\bibitem[Chang \& Refsdal(1979)]{1979Natur.282..561C} Chang, K., \& Refsdal, S.\ 1979, \nat, 282, 561 
\bibitem[Chang \& Refsdal(1984)]{1984A&A...132..168C} Chang, K., \& Refsdal, S.\ 1984, \aap, 132, 168
\bibitem[Eigenbrod et al.(2008a)]{2008A&A...480..647E} Eigenbrod, A., Courbin, F., Sluse, D., Meylan, G., \& Agol, E.\ 2008a, \aap, 480, 647 
\bibitem[Eigenbrod et al.(2008b)]{2008A&A...490..933E} Eigenbrod, A., Courbin, F., Meylan, G., et al.\ 2008b, \aap, 490, 933 
\bibitem[Gil-Merino et al.(2005)]{2005A&A...432...83G} Gil-Merino, R., Wambsganss, J., Goicoechea, L.~J., \& Lewis, G.~F.\ 2005, \aap, 432, 83 
\bibitem[Gil-Merino et al.(2006)]{2006MNRAS.371.1478G} Gil-Merino, R., Gonz{\'a}lez-Cadelo, J., Goicoechea, L.~J., Shalyapin, V.~N., \& Lewis, G.~F.\ 2006, \mnras, 371, 1478 
\bibitem[Goicoechea et al.(2003)]{2003A&A...397..517G} Goicoechea, L.~J., Alcalde, D., Mediavilla, E., \& Mu{\~n}oz, J.~A.\ 2003, \aap, 397, 517 
\bibitem[Kochanek(2004)]{2004ApJ...605...58K} Kochanek, C.~S.\ 2004, \apj, 605, 58
\bibitem[Koptelova et al.(2007)]{2007MNRAS.381.1655K} Koptelova, E., Shimanovskaya, E., Artamonov, B., \& Yagola, A.\ 2007, \mnras, 381, 1655 
\bibitem[Marshall et al.(2010)]{2010AAS...21540115M} Marshall, P.~J., Bradac, M., Chartas, G., et al.\ 2010, Bulletin of the American Astronomical Society, 42, \#401.15 
\bibitem[Mediavilla et al.(2006)]{2006ApJ...653..942M} Mediavilla, E., Mu{\~n}oz, J.~A., Lopez, P., et al.\ 2006, \apj, 653, 942 
\bibitem[Mediavilla et al.(2011)]{2011ApJ...741...42M} Mediavilla, E., 
Mediavilla, T., Mu{\~n}oz, J.~A., et al.\ 2011, \apj, 741, 42 
\bibitem[Mediavilla et al.(2015)]{2015ApJ...798..138M} Mediavilla, E., Jimenez-Vicente, J., Mu{\~n}oz, J.~A., Mediavilla, T., \& Ariza, O.\ 2015, \apj, 798, 138 
\bibitem[Moriyama \& Mineshige(2015)]{2015PASJ..tmp..240M} Moriyama, K., \& Mineshige, S.\ 2015, \pasj, 240
\bibitem[Mosquera et al.(2009)]{2009ApJ...691.1292M} Mosquera, A.~M., Mu{\~n}oz, J.~A., \& Mediavilla, E.\ 2009, \apj, 691, 1292 
\bibitem[Mosquera \& Kochanek(2011)]{2011ApJ...738...96M} Mosquera, A.~M., \& Kochanek, C.~S.\ 2011, \apj, 738, 96 
\bibitem[Mu\~{n}oz et al.(2015)]{} Mu\~{n}oz, J. A., Vives-Arias, H., Mosquera, A. M., Jimenez-Vicente, J., Kochanek, C. S., Mediavilla, E.,  2015, \apj, submitted, arXiv:1509.04305.
\bibitem[Poindexter \& Kochanek(2010)]{Poindexter2010} Poindexter, S., \& Kochanek, C.~S.\ 2010, \apj, 712, 668
\bibitem[Shakura \& Sunyaev(1973)]{1973A&A....24..337S} Shakura, N.~I., \& Sunyaev, R.~A.\ 1973, \aap, 24, 337 
\bibitem[Shalyapin(2001)]{2001AstL...27..150S} Shalyapin, V.~N.\ 2001, Astronomy Letters, 27, 150 
\bibitem[Shalyapin et al.(2002)]{2002ApJ...579..127S} Shalyapin, V.~N., Goicoechea, L.~J., Alcalde, D., et al.\ 2002, \apj, 579, 127
\bibitem[Schneider \& Weiss(1987)]{1987A&A...171...49S} Schneider, P., \& Weiss, A.\ 1987, \aap, 171, 49 
\bibitem[Sluse et al.(2011)]{2011A&A...528A.100S} Sluse, D., Schmidt, R., Courbin, F., et al.\ 2011, \aap, 528, A100 
\bibitem[Udalski et al.(2006)]{2006AcA....56..293U} Udalski, A., Szymanski, 
M.~K., Kubiak, M., et al.\ 2006, \actaa, 56, 293 
\bibitem[Wambsganss(2006)]{2006glsw.conf..453W} Wambsganss, J.\ 2006, Saas-Fee Advanced Course 33: Gravitational Lensing: Strong, Weak and Micro, 453
\bibitem[Wo{\'z}niak et al.(2000)]{2000ApJ...529...88W} Wo{\'z}niak, P.~R., Alard, C., Udalski, A., et al.\ 2000, \apj, 529, 88 
\bibitem[Wyithe et al.(1999)]{1999MNRAS.309..261W} Wyithe, J.~S.~B., Webster, R.~L., \& Turner, E.~L.\ 1999, \mnras, 309, 261 
\bibitem[Wyithe et al.(2000)]{2000MNRAS.318.1120W} Wyithe, J.~S.~B., Webster, R.~L., \& Turner, E.~L.\ 2000, \mnras, 318, 1120 
\end{thebibliography}
\end{document}